\UseRawInputEncoding
\documentclass[preprint,prc,amsmath,amssymb,superscriptaddress,noffotinbib,preprintnumbers]{revtex4-1}
\usepackage{url,color,colordvi,epsfig,pstricks}
\usepackage{hyperref}
\usepackage{cleveref}
\usepackage{graphicx,bm,dcolumn}
\hypersetup{
    colorlinks=true, 
    linktoc=all, 
    linkcolor=blue, 
    citecolor=blue}

\begin{document}

\title{Cross sections for coherent elastic and inelastic neutrino-nucleus scattering}

\author{N.~Van Dessel}
\affiliation{Department of Physics and Astronomy, Ghent University, Proeftuinstraat 86, B-9000 Gent, Belgium}
\author{V.~Pandey}\email{vpandey@fnal.gov}
\affiliation{Fermi National Accelerator Laboratory, Batavia, Illinois 60510, USA}
\author{H.~Ray}
\affiliation{Department of Physics, University of Florida, Gainesville, FL 32611, USA}
\author{N.~Jachowicz}\email{natalie.jachowicz@ugent.be}
\affiliation{Department of Physics and Astronomy, Ghent University, Proeftuinstraat 86, B-9000 Gent, Belgium}

\begin{abstract}

The prospects of extracting new physics signals in  coherent elastic neutrino--nucleus scattering (CE$\nu$NS) processes are limited by the precision with which the underlying nuclear structure physics, embedded in the weak nuclear form factor, is known. We present calculations of charge and weak nuclear form factors and CE$\nu$NS cross sections on $^{12}$C, $^{16}$O, $^{40}$Ar, $^{56}$Fe and $^{208}$Pb nuclei. We obtain the proton and neutron densities, and charge and weak form factors by solving Hartree--Fock (HF) equations with a Skyrme (SkE2) nuclear potential. We validate our approach by comparing $^{208}$Pb and $^{40}$Ar charge form factor predictions with available elastic electron scattering data.
Since CE$\nu$NS experiments at stopped--pion sources are also well suited to measure  inelastic charged--current and neutral--current neutrino--nucleus cross sections, we also present calculations for these processes, incorporating a continuum Random Phase Approximation (CRPA) description on top of the HF-SkE2 picture of the nucleus.  Providing both coherent as well as inelastic cross sections in a consistent framework, we aim at obtaining a reliable and detailed comparison of the strength of these processes in the energy region below ~100 MeV.
Furthermore, we attempt to gauge the level of theoretical uncertainty pertaining to the description of the  $^{40}$Ar form factor and CE$\nu$NS cross sections by comparing relative differences between recent microscopic nuclear theory and widely--used phenomenological form factor predictions. Future precision measurements of CE$\nu$NS will potentially help in constraining these nuclear structure details that will in turn improve prospects of extracting new physics.

\end{abstract}

\maketitle


\section{Introduction}

\label{Sec:Introduction}

\par Coherent elastic neutrino--nucleus scattering (CE$\nu$NS), where the only detectable reaction product is a low momentum recoiling nucleus, was suggested soon after the experimental discovery of a weak neutral current in neutrino interactions~\cite{Freedman:1974}. Even though for neutrino energies of some tens of MeV the CE$\nu$NS cross section is a few orders of magnitude larger than competing inelastic processes, the difficulty in detecting the $\sim$keV scale recoil of a nucleus has hindered experimental detection of this process for decades. In 2017, the COHERENT collaboration detected the first CE$\nu$NS signal using a stopped--pion beam in the Spallation Neutron Source (SNS) at Oak Ridge National Laboratory with a CsI detector ~\cite{COHERENT:2017, COHERENT:2018}, followed up by another recent measurements in a liquid argon (LAr) detector~\cite{COHERENT:2019, COHERENT:2020} and in CsI detector~\cite{COHERENT:2021}.

\par The detection of CE$\nu$NS has opened up a slew of opportunities in high--energy physics, astrophysics and in nuclear physics, inspiring new probes into beyond--Standard--Model (BSM) physics and new experimental methods. Several extensions of the SM that can be explored at low energy such as non--standard interactions (NSI)~\cite{Liao:2017, Dent:2018, Aristizabal:2018, Denton:2018}, sterile neutrinos~\cite{Kosmas:2017, Blanco:2019}, CP--violation~\cite{Aristizabal:2019_a}, as well as exploration of nuclear effects, are being studied~\cite{Cadeddu:2018, Ciuffoli:2018, Aristizabal:2019, Papoulias:2019}. Several experimental programs have been or are being set up to detect CE$\nu$NS and BSM signals in the near future using stopped--pion neutrino sources in COHERENT at the SNS~\cite{COHERENT:2017}, Coherent CAPTAIN--Mills (CCM) at Los Alamos National Laboratory (LANL)~\cite{CCM} and at the proposed European Spallation Source (ESS) facility~\cite{ESS}, as well as reactor--produced neutrinos in CONNIE~\cite{CONNIE}, MINER~\cite{MINER}, $\nu$GEN~\cite{vGEN}, NUCLEUS~\cite{NUCLEUS}, RICOCHET~\cite{RICOCHET}, TEXONO~\cite{TEXONO}, NEON~\cite{NEON} and vIOLETA~\cite{vIOLETA}. 

\par The main source of uncertainty in the evaluation of the CE$\nu$NS cross section is the accuracy with which the underlying nuclear structure and nucleon dynamics that determine the distributions of the nucleon density in the nuclear ground state, embedded in the form factor, are known in the target nucleus. The ground state proton (charge) density distributions are relatively well constrained through elastic electron scattering experiments pioneered by Hofstadter and collaborators at the Stanford Linear Accelerator~\cite{Hofstadter:1956}, followed by other measurements in the following decades~\cite{Vries:1987,Fricke:1995, Angeli:2013}. CE$\nu$NS is however primarily sensitive to the neutron density distributions of the nucleus, which are only poorly constrained. Hadronic probes have been used to extract neutron distributions,  these measurements are however plagued by ill--controlled model--dependent uncertainties associated with the strong interaction~\cite{Thiel:2019}. More (experimentally) challenging electroweak probes such as parity--violating electron scattering (PVES)~\cite{Donnelly:1989, Thiel:2019} and CE$\nu$NS provide relatively model--independent ways of determining neutron distributions. In recent years, one such PVES experiment, PREX at Jefferson lab, has measured the weak charge of $^{208}$Pb at a single value of momentum transfer~\cite{PREX:2012, Horowitz:2012}, while a follow up PREX--II experiment is ongoing to improve the precision of that measurement. Another PVES experiment, CREX at Jefferson lab, is underway to measure the weak form factor of $^{48}$Ca~\cite{CREX}. Future ton and multi-ton CE$\nu$NS detectors will enable more precise measurements and will potentially offer a powerful avenue to constrain neutron density distributions and weak form factors of nuclei at low momentum transfers where the process remains coherent~\cite{Cadeddu:2018, Ciuffoli:2018, Patton:2012}.

\par As long as no precision measurements of neutron density distributions of nuclei are available, the weak nuclear form factor has to be modeled in order to evaluate the CE$\nu$NS cross section and event rates. The accuracy of such an assumption is vital to the CE$\nu$NS program since any experimentally measured deviation from the expected CE$\nu$NS event rate can point to new physics or to unconstrained nuclear physics. It is therefore crucial to treat the underlying nuclear structure physics that is embedded in nuclear form factors with utmost care. Phenomenological approaches, such as the Klein--Nystrand form factor~\cite{KN:1999} adapted by the COHERENT collaboration, or the Helm form factor~\cite{Helm:1956} where density distributions are represented by analytical expressions, are widely used in the CE$\nu$NS community. Empirical values of the proton rms radius, measured in elastic electron scattering, are often used to evaluate the proton form factor and often similar parameterizations are assumed for the neutron form factor. Microscopic nuclear physics approaches which provide a more accurate description of the nuclear ground state and density distributions such as density functional theory~\cite{Patton:2012}, coupled--cluster theory from first principles~\cite{Payne:2019}, relativistic mean--field model~\cite{Yang:2019}, Hartree--Fock plus Bardeen--Cooper--Schrieffer model~\cite{Co:2020} as well as effective field theory approaches~\cite{Hoferichter:2020, Tomalak:2020} have also been reported in recent years.

\par In this work we will present a microscopic many--body nuclear theory model where the nuclear ground state is described in a Hartree--Fock (HF) approach with a Skyrme (SkE2) nuclear potential.  We calculate proton and neutron density distributions, charge and weak form factors, and CE$\nu$NS cross sections on $^{12}$C, $^{16}$O, $^{40}$Ar, $^{56}$Fe and $^{208}$Pb, and confront our predictions with the available experimental data. In view of the  worldwide interest in liquid--argon--based neutrino and dark matter experiments, we pay special attention to the $^{40}$Ar nucleus. We attempt to gauge the level of theoretical uncertainty pertaining to the description of the $^{40}$Ar form factor and CE$\nu$NS cross section by comparing relative differences between recent nuclear theory and widely--used phenomenological form factor predictions. 

\par CE$\nu$NS experiments at stopped--pion sources are also well-suited to measure inelastic neutrino--nucleus cross sections. These measurements, in particular on $^{40}$Ar, will provide powerful constraints on supernova detection capabilities of future kiloton neutrino experiments. To this end, we also present inelastic charged--current (CC) and neutral--current (NC) cross section calculations on $^{40}$Ar, incorporating a continuum Random Phase Approximation (CRPA) description on top of the initial HF--SkE2 picture of the nucleus.

\par The remainder of this manuscript is organized as follows. In Sec.~\ref{Sec:Formalism}, we lay out the general formalism of calculating the CE$\nu$NS and inelastic neutrino--nucleus scattering cross section. In Sec.~\ref{Sec:Results}, we present results of proton and neutron densities, charge and weak form factors, and CE$\nu$NS cross sections on $^{12}$C, $^{16}$O, $^{40}$Ar, $^{56}$Fe and $^{208}$Pb obtained within our HF--SkE2 approach. We focus on $^{40}$Ar in subsection~\ref{Subsec:Argon}, and compare our predictions with experimental data and other theoretical calculations. We also present inelastic cross sections on $^{40}$Ar in subsection~\ref{Subsec:Argon}. We present conclusions of this study in Sec.~\ref{Sec:Conclusions}.


\section{Formalism}
\label{Sec:Formalism}

\par In this section, we lay out the general formalism for calculating cross sections of the coherent elastic and inelastic neutrino-nucleus scattering process. 


\subsection{CE$\nu$NS Cross Section}
\label{Subsec:CEvNS_CS}

\par
\begin{figure}
\centering
\includegraphics[width=0.44\columnwidth]{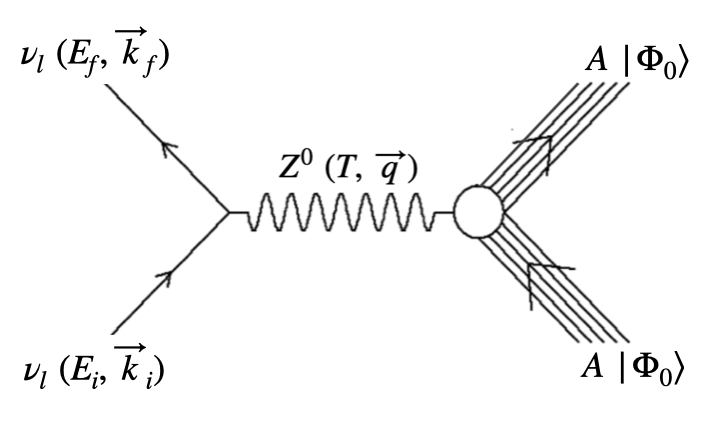}
\includegraphics[width=0.45\columnwidth]{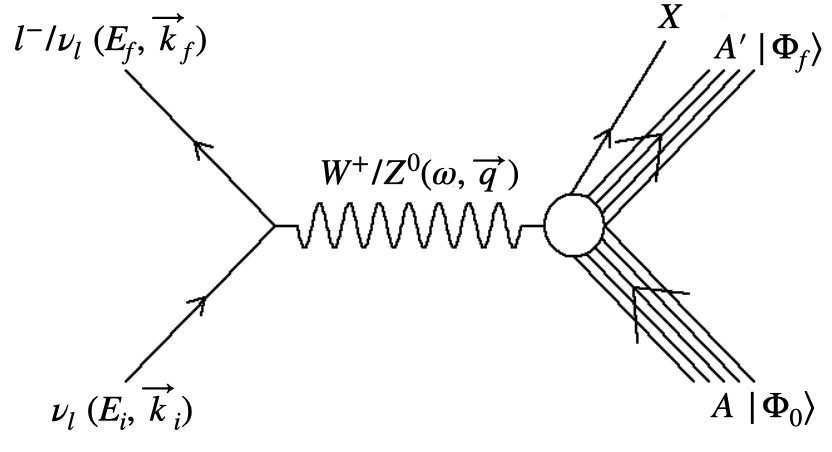}
\caption{(Left) Diagrammatic representation of the CE$\nu$NS process where a single $Z^{0}$ boson is exchanged between neutrino and target nucleus. (Right) Diagrammatic representation of the inelastic neutrino-nucleus scattering where a single $W^+$ (CC) or $Z^{0}$ (NC) boson is exchanged between neutrino and target nucleus.}
\label{Fig:scattering_diagram}
\end{figure}

 A neutrino with four momentum $k_i = (E_i,\vec{k}_i)$ scatters off the nucleus, which is initially at rest in the lab frame with $p_{A} = (M_A,\vec{0})$, exchanging a $Z^{0}$ boson. The neutrino scatters off, carrying away four momentum $k_f = (E_f,\vec{k}_f)$ while the nucleus remains in its ground state and receives a small recoil energy $T$, so that $p'_{A} = (M_A + T,\vec{p}'_{A})$ with $|\vec{p}'_{A}| = \sqrt{(M_A + T)^2 - M_A^2}$ and $T = q^2/2M_{A}$. Here, $M_A$ is the rest mass of the nucleus, $q = |\vec{q}|$ is the absolute value of the three--momentum transfer which is of the order of keV for neutrino energies of tens of MeV, $Q^2 \approx q^2 = |\vec{k}_f-\vec{k}_i|^2$, and the velocity dependent factor in the denominator refers to the relative velocity of the interacting particles. The process is schematically shown in Fig.~\ref{Fig:scattering_diagram}.

\par The initial elementary expression for the cross section reads
\begin{equation}
\begin{aligned}
\mathrm{d}^6\sigma &=\frac{1}{\left| \vec{v}_i - \vec{v}_A \right|}\frac{m_i}{E_i}\frac{m_f}{E_f}\frac{\mathrm{d}^3\vec{k}_f}{(2\pi)^3}\frac{M_A}{M_A + T}\frac{\mathrm{d}^3\vec{p}'_{A}}{(2\pi)^3} \\
 &\times (2\pi)^4  \overline{\sum}_{fi}\left| \mathcal{M} \right|^2 \delta^{(4)}(k_i + p_A - k_f -p'_A).
\end{aligned}
\end{equation}
This expression can be integrated  to yield the expression for the cross section differential in neutrino scattering angle $\theta_f$:
\begin{equation}
\begin{aligned}
\frac{\mathrm{d}\sigma}{ \mathrm{d}\cos{\theta_f}} &=\frac{m_i}{E_i}\frac{m_f}{E_f}\frac{M_A}{M_A + T} \frac{E_f^2}{2\pi}f_{rec}^{-1} \overline{\sum}_{fi}\left| \mathcal{M} \right|^2.
\end{aligned}
\end{equation}
The recoil factor reads
\begin{equation}
f_{rec} = \frac{E_i}{E_f}\frac{M_A}{M_A+T}.
\end{equation}
Working out the Feynman amplitude one gets
\begin{equation}
\overline{\sum}_{fi}\left| \mathcal{M} \right|^2 = \frac{G_F^2}{2}L_{\mu\nu}W^{\mu\nu},
\end{equation}
with the nuclear tensor $W^{\mu\nu}$ reading
\begin{equation}
W^{\mu\nu} = \overline{\sum}_{fi} (\mathcal{J}^{\mu}_{nuc})^\dagger \mathcal{J}^{\nu}_{nuc}.
\end{equation}
The summation symbols in these expressions denote summing and averaging over initial and final polarizations respectively. The nuclear tensor depends on the nuclear current transition amplitudes:
\begin{equation}
\mathcal{J}^{\mu}_{nuc} = \langle \Phi_\textrm{0} | \widehat{J}^\mu(\vec{q}) | \Phi_\textrm{0} \rangle .
\end{equation}

\par Under the assumption that the nuclei of interest are spherically symmetric with $J^\pi = 0^+$ and taking the z--axis to be along the direction of $\vec{q}$, one only needs to take into account the zeroth and third component of the nuclear current's vector part, which are furthermore connected through vector current conservation (CVC):
\begin{equation}
q^\mu \widehat{J}_\mu(\vec{q}) = 0.
\end{equation}

\par Through performing the necessary algebra, one arrives at the final expression
\begin{equation}
\frac{\mathrm{d}\sigma}{ \mathrm{d}\cos{\theta_f}} = \frac{G_F^2}{2\pi} \frac{E_f^3}{E_i} \left[\frac{Q^4}{q^4}(1+\cos{\theta_f}) |\mathcal{J}^V_0|^2\right]
\end{equation}
where $\mathcal{J}^V_0$ is the transition amplitude induced by the nuclear current. One can then safely approximate $\frac{Q^4}{q^4} \approx 1$ and express the differential cross section as a function of the neutrino scattering angle $\theta_f$ as:
\begin{equation}\label{Eq:xs_angular}
\frac{\mathrm{d}\sigma}{ \mathrm{d}\cos{\theta_f}} = \frac{G_F^2}{2\pi}  \frac{E_f^3}{E_i} (1+\cos{\theta_f}) \frac{Q_W^2}{4}F_{W}^2(Q^2)
\end{equation}
where $G_F$ is the Fermi coupling constant, and $Q_W$ the weak nuclear charge :
\begin{equation}\label{eq:weakcharge}
Q^{2}_{W} = [g_p^V Z+g_n^V N]^2 = [(1-4\sin^2\theta_\text{W}) Z-N]^2
\end{equation}
with coupling constants $g_n^V = -1$ and $g_p^V = (1-4\sin^2\theta_\text{W})$. $N$ and $Z$ are the nucleus' neutron and proton number, and $\theta_W$ is the weak mixing angle. The value is such that $\sin^2{\theta_W} = 0.23857$, which is valid at low momentum transfers~\cite{PDG2018}. \\

Here we have introduced the elastic form factor, $F_{W}^2(Q^2)$, which we will discuss later in this subsection. In elastic scattering the entire nuclear dynamics is encoded in this form factor. Equivalently one can express the differential cross section as a function of the nuclear recoil $T$, which reads:
\begin{equation}\label{Eq:xs_recoil}
\frac{\mathrm{d}\sigma}{ \mathrm{d}T} = \frac{G^{2}_{F}}{\pi} M_{A} \left(1-\frac{T}{E_{i}}-\frac{M_A T}{2 E^2_i}\right)~\frac{Q^2_{W}}{4}~F_{W}^2(Q^2), 
\end{equation}

\par In Eq.~(\ref{Eq:xs_angular}) and~(\ref{Eq:xs_recoil}), we have expressed the CE$\nu$NS kinematic distribution both in neutrino scattering angle, $\theta_f$, and in nuclear recoil energy $T$. In most experiments the only signal of a CE$\nu$NS event is a nuclear recoil energy deposition. In principle, future experiments with more advanced detector technologies may be able to detect both nuclear recoil and angular distribution simultaneously. Such capabilities are already being explored in some dark-matter experiments and will greatly enhance the physics capabilities of future CE$\nu$NS experiments~\cite{Abdullah:2020}. 

\par The scattering process' cross section is proportional to the squared magnitude of the transition amplitude induced by the nuclear current. Since the relevant ground state to ground state transition for spherically symmetrical nuclei is $0^+ \rightarrow 0^+$, only the vector part of the current will contribute. The amplitude can be expressed as
\begin{equation}\label{Eq:cevnsformfactor1}
\begin{aligned}
 \mathcal{J}^V_0 &= \langle \Phi_0 | \widehat{J}_0^V(\vec{q}) | \Phi_0 \rangle \\
 &= \int e^{i\vec{q}\cdot \vec{r}} \langle \Phi_0 | \widehat{J}_0^V(\vec{r}) | \Phi_0 \rangle \\
 &= \frac{1}{2}\left[\left(1 - 4 \sin^2{\theta_W} \right) f_p(\vec{q})F_{p}(Q^2) \right. \\
 &- \left. f_n(\vec{q})F_{n}(Q^2) \right],
\end{aligned}
\end{equation}
where we have inserted the impulse approximation (IA) expression for the nuclear current, as a sum of single--body operators:
\begin{equation}\label{Eq:cevnscurrent}
\widehat{J}_0^V(\vec{r}) = \sum_i F^Z(Q^2,i)\delta^{(3)}(\vec{r}-\vec{r}_i), 
\end{equation}
with
\begin{equation}\label{Eq:cevnscurrent2}
\begin{aligned}
F^Z(Q^2,i) &= \left( \frac{1}{2}-\sin^2{\theta_W} \right)(F_{p} - F_{n})\tau_3(i) \\
&- \sin^2{\theta_W}(F_{p} + F_{n}),
\end{aligned}
\end{equation}
where we used the convention $\tau_3(i)=+1$ for proton, -1 for neutrons.
 Furthermore, $f_p(\vec{q})$ and $f_n(\vec{q})$ are the Fourier transforms of the proton and neutron densities, respectively. $F_{p}$ and $F_{n}$ are proton and neutron form factors, for which we adopt the standard Galster parametrization. Note that using a more sophisticated parametrization of the form factor, other than Galster, will not affect the results at the energies relevant to this work.
 The overall structure of the transition amplitude consists of products of the weak charge with two factors: the nuclear form factor, determined by the spatial distribution of the nucleons in the nucleus, as well as the nucleon form factor. We arrive at the expression:
\begin{equation}\label{Eq:cevnsformfactor2}
\begin{aligned}
 F_{W}(Q^2) &= \frac{1}{Q_W}\left[\left(1 - 4 \sin^2{\theta_W} \right) f_p(\vec{q})F_{p}(Q^2) \right. \\
  & \left. -  f_n(\vec{q})F_{n}(Q^2) \right] = \frac{2}{Q_W}\mathcal{J}^V_0,
 \end{aligned}
\end{equation}
 such that the form factor becomes 1 in the static limit. Note that in writing down the functional dependence we can make use of the non--relativistic approximation $Q \approx |\vec{q}|$, valid in the energy regime considered. 

\par We employ a microscopic many--body nuclear theory model where the nuclear ground state is described in a Hartree--Fock (HF) approach with a Skyrme (SkE2) nuclear potential, which we will refer to as HF--SkE2. We solve the HF equations to obtain single--nucleon wave functions for the bound nucleons in the nuclear ground state. We evaluate proton ($\rho_{p}(r)$) and neutron ($\rho_{n}(r)$) density distributions from those wave functions. The proton density is utilized to calculate charge form factor (which can also be referred to as electromagnetic form factor), $F_{ch}(Q^2)$, while both proton and neutron densities are utilized to compute weak, $F_W(Q^2)$, nuclear form factor, as shown in Eq.~(\ref{Eq:cevnsformfactor2}). This approach involves more realistic nuclear structure calculations of proton and neutron density distributions making it more reliable compared to the phenomenological approaches that rely on the approximation $\rho_{n}(r) \approx \rho_{p}(r)$, utilizing empirical values of $\rho_{p}(r)$ extracted from electron scattering experiments. 


{\renewcommand{\arraystretch}{1.2}}
\begin{table}
   \centering
\begin{tabular}{|l|l|l|l|l|l|}
  \hline
   $ p/n $ & $i$ & $n_i, l_i, j_i $ & $\varepsilon_i$ (MeV) & $v_i^2$ & \# N \\
   \hline
   p & 1 & $1s_{1/2}$ & -43.7029 & 1.00 & 2 \\
   \hline
   p & 2 & $1p_{3/2}$ & -31.4496 & 1.00 & 4 \\
   \hline
   p & 3 & $1p_{1/2}$ & -27.3921 & 1.00 & 2 \\
   \hline
   p & 4 & $1d_{5/2}$ & -17.7027 & 1.00 & 6 \\
   \hline
   p & 5 & $2s_{1/2}$ & -12.0822 & 1.00 & 2 \\
   \hline
   p & 6 & $1d_{3/2}$ & -10.9243 & 0.50 & 2 \\
   \hline
   n & 1 & $1s_{1/2}$ & -48.3047 & 1.00 & 2 \\
   \hline
   n & 2 & $1p_{3/2}$ & -35.2020 & 1.00 & 4 \\
   \hline
   n & 3 & $1p_{1/2}$ & -31.0247 & 1.00 & 2 \\
   \hline
   n & 4 & $1d_{5/2}$ & -21.1035 & 1.00 & 6 \\
   \hline
   n & 5 & $2s_{1/2}$ & -16.1116 & 1.00 & 2 \\
   \hline
   n & 6 & $1d_{3/2}$ & -14.0266 & 1.00 & 4 \\
   \hline
   n & 7 & $1f_{7/2}$ & -7.2108 & 0.25 & 2 \\
   \hline
\end{tabular}
\caption{Single--particle energies in $^{40}$Ar, as provided by a HF calculation using the SkE2 interaction.}
\label{table:arlevels}
\end{table} 

\subsection{Inelastic Cross Sections}
\label{Subsec:crpaqe}

\par CE$\nu$NS experiments at stopped--pion sources are also sensitive to inelastic neutrino--nucleus interactions. In several astrophysical environments elastic and inelastic processes come in competition.  To this end, we also present calculations of inelastic charged--current (CC) and neutral--current (NC) cross sections, calculated within the same framework. These results are obtained by including effects of long-range correlations through a continuum Random Phase Approximation (CRPA) description on top of the HF--SkE2 initial picture of the nucleus. 

\par The inelastic neutrino--nucleus scattering process is schematically shown in Fig.~\ref{Fig:scattering_diagram}. A neutrino with four momentum $k_i = (E_i,\vec{k}_i)$ scatters off the nucleus, which is initially at rest in the lab frame, exchanging a $W^+$ (CC) or a $Z^{0}$ (NC) boson. The nucleus receives four momentum $Q = (\omega, \vec{q})$, where $\omega = E_i - E_f$ and $\vec{q} = \vec{k}_i - \vec{k}_f$, while the scattered lepton carries away four momentum $k_f = (E_f,\vec{k}_f)$. Since we concern ourselves with inclusive calculations, the hadronic part of the final states are integrated out. The inelastic neutrino--nucleus differential cross section of this process can be written as 
\begin{equation}\label{eq:xsec}
\begin{aligned}
\frac{\mathrm{d}^3\sigma}{\mathrm{d}\omega\mathrm{d}\Omega} =& \sigma_W E_f k_f \zeta^2(Z',E_f) \\
&\times \left( v_{CC} R_{CC} + v_{CL} R_{CL} + v_{LL} R_{LL} \right.  \\
& + \left. v_{T} R_{T} + h v_{T'} R_{T’} \right),
\end{aligned}
\end{equation}
with the Mott-like cross section prefactor $\sigma_W$ defined as
\begin{equation*}
\sigma_W^{CC} = \left(\frac{G_F \cos{\theta_c}}{2\pi} \right)^2, ~\sigma_W^{NC} = \left(\frac{G_F}{2\pi} \right)^2,
\end{equation*}
where $G_F$ is the Fermi constant and $\cos{\theta_c}$ the Cabibbo angle. The factor $\zeta^2(Z',E_f)$ is introduced in order to take into account the distortion of the scattered lepton wave function in the Coulomb field of the final nucleus with $Z'$ protons, in the case of CC interaction~\cite{VanDessel:2019}. In the NC case $\zeta^2(Z,E_f)$ equals $1$. The influence of the lepton helicity on the cross section is encoded in $h$ which is + for neutrinos and −- for antineutrinos. 

The $v$--factors are leptonic functions that are entirely determined by lepton kinematics. The $R$--factors are the nuclear response functions that depend on the energy and momentum transfer ($\omega$, $q$) and contain all the nuclear information involved in this process. The indices $L$ and $T$ correspond to longitudinal and transverse contributions, relative to the direction of the momentum transfer. The nuclear responses are function of the transition amplitude, ${J}_\mu^{nucl}(\omega,q)$, between the initial $| \Phi_\textrm{0} \rangle$ and final $| \Phi_\textrm{f} \rangle$ state:
\begin{equation}\label{eq:current}
{J}_\mu^{nucl}(\omega,q) = \langle \Phi_\textrm{f} | \hat{J}_\mu(q) | \Phi_\textrm{0} \rangle,
\end{equation}
where the nuclear current, $\hat{J}_\mu({q})$, is the Fourier transform of the nuclear current operator in coordinate space:
\begin{equation}
\hat{J}_\mu(q) = \int \mathrm{d} {x} e^{i{x}\cdot{q}} \hat{J}_\mu({x}).
\end{equation}

These are computed within a HF-CRPA framework. For a detailed discussion of the nuclear response we refer the reader to our previous work in Refs.~\cite{Ryckebusch:1988, Ryckebusch:1989, Jachowicz:1999, Jachowicz:2002, Jachowicz:2002_2, Jachowicz:2004, Jachowicz:2006, Pandey:2014, Pandey:2015, Pandey:2016, VanDessel:2018, VanDessel:2019, Nikolakopoulos:2019, VanDessel:2019_2, Nikolakopoulos:2020}. Here we briefly describe the essence of our approach. The CRPA description goes beyond a pure spectator approach, incorporating long--range correlations in the cross section calculations. Within many--body theory, the random phase approximation achieves this by modeling excitations as superpositions of particle-hole ($ph^{-1}$) and hole-particle ($hp^{-1}$) states out of a correlated ground state:
\begin{equation}\label{eq:tradrpa}
| \Psi^C_{RPA} \rangle = \sum_{C'} \left\lbrace X_{c,c'} |p'h'^{-1}\rangle - Y_{c,c'} |h'p'^{-1}\rangle \right\rbrace,
\end{equation}
where the summation index $C$ denotes a set of quantum numbers defining an excitation channel unambiguously:
\begin{equation}
C = \left\lbrace n_h,l_h,j_h,m_{j_h},\varepsilon_h;l_p,j_p,m_{j_p},\tau_z \right\rbrace.
\end{equation}
The indices $p$ and $h$ represent the quantum numbers related to the particle or the hole state, $\varepsilon_h$ denotes the binding-energy of the hole state and $\tau_z$ defines the isospin character of the particle-hole pair. Since the RPA approach describes nuclear excitations as the coherent superposition of individual particle-hole states out of a correlated ground state, it allows the description of collective effects in the nucleus. 

\begin{figure*}
\centering
\includegraphics[width=0.9\textwidth]{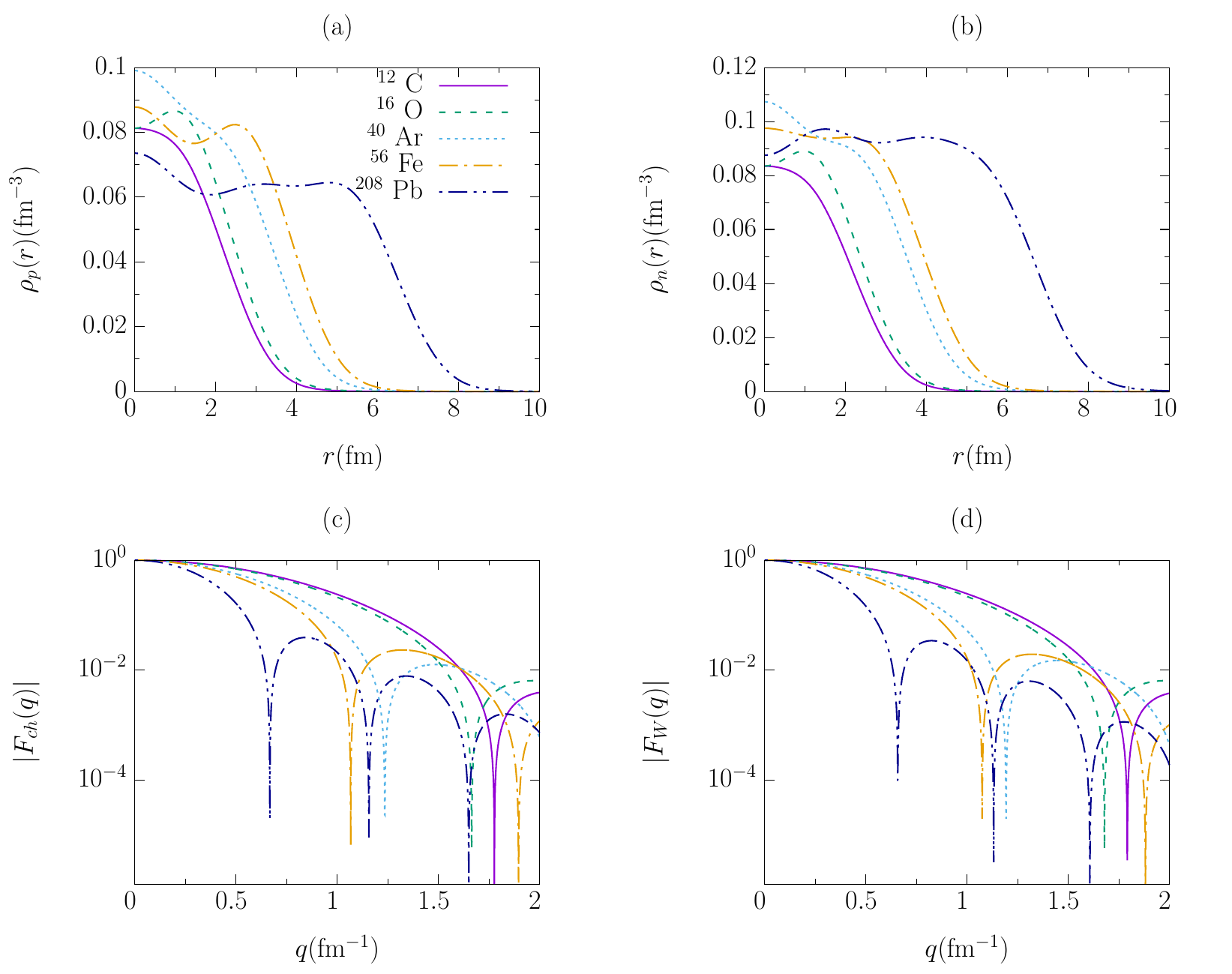}
\caption{(color online) Panels (a) and (b) represent proton and neutron densities of different nuclei obtained using the HF--SkE2 approach. Panels (c) through (d) represent charge and weak form factors for the different nuclei.}
\label{Fig:densandform}
\end{figure*}

Besides Eq.~(\ref{eq:tradrpa}), the RPA approach can also be formulated in a propagator description of many--body theory, where the central object containing the information on the excited states of the many--body system is the  polarization propagator. In the Lehmann representation, the CRPA approach involves solving the RPA equation for the local polarization propagator $\Pi^{RPA}(x_1,x_2,E_{exc})$ in coordinate space:
\begin{equation}\label{eq:locpolproprpa}
\begin{aligned}
&\Pi^{RPA}(x_1,x_2,E_{exc}) = \Pi^{(0)}(x_1,x_2,E_{exc}) \\ 
&+ \frac{1}{\hbar} \int \mathrm{d}x \int \mathrm{d}x' \left[ \Pi^{(0)}(x_1,x,E_{exc}) \right. \\ &\times \left. \tilde{V}(x,x') \Pi^{RPA}(x',x_2,E_{exc}) \right],
\end{aligned}
\end{equation}
where $E_{exc}$ is the excitation energy of the target nucleus and $x$ is the shorthand notation for the combination of the spatial, spin, and isospin coordinates. In this equation, the antisymmetrized residual interaction $\tilde{V}(x,x')$, is the same SkE2 Skyrme interaction we have utilized to calculate the single particle wave functions (and therefore, nuclear densities) of the CE$\nu$NS cross sections, keeping the scheme self–consistent. $\Pi^{(0)}(x_1,x_2,E_{exc})$ denotes the zeroth-order contribution to the polarization propagator which is equivalent to the HF contribution. The (local) polarization propagator $\Pi^{RPA}(x_1,x_2,E_{exc})$, which describes the propagation of particle–hole pairs, is obtained by adding the iteration of first-order contributions to the bare local polarization propagator $\Pi^{(0)}(x_1,x_2,E_{exc})$. By solving this equation, one obtains the CRPA transition amplitudes needed to calculate the inelastic neutrino--nucleus cross sections. 

\begin{figure}
\centering
\includegraphics[width=0.7\columnwidth]{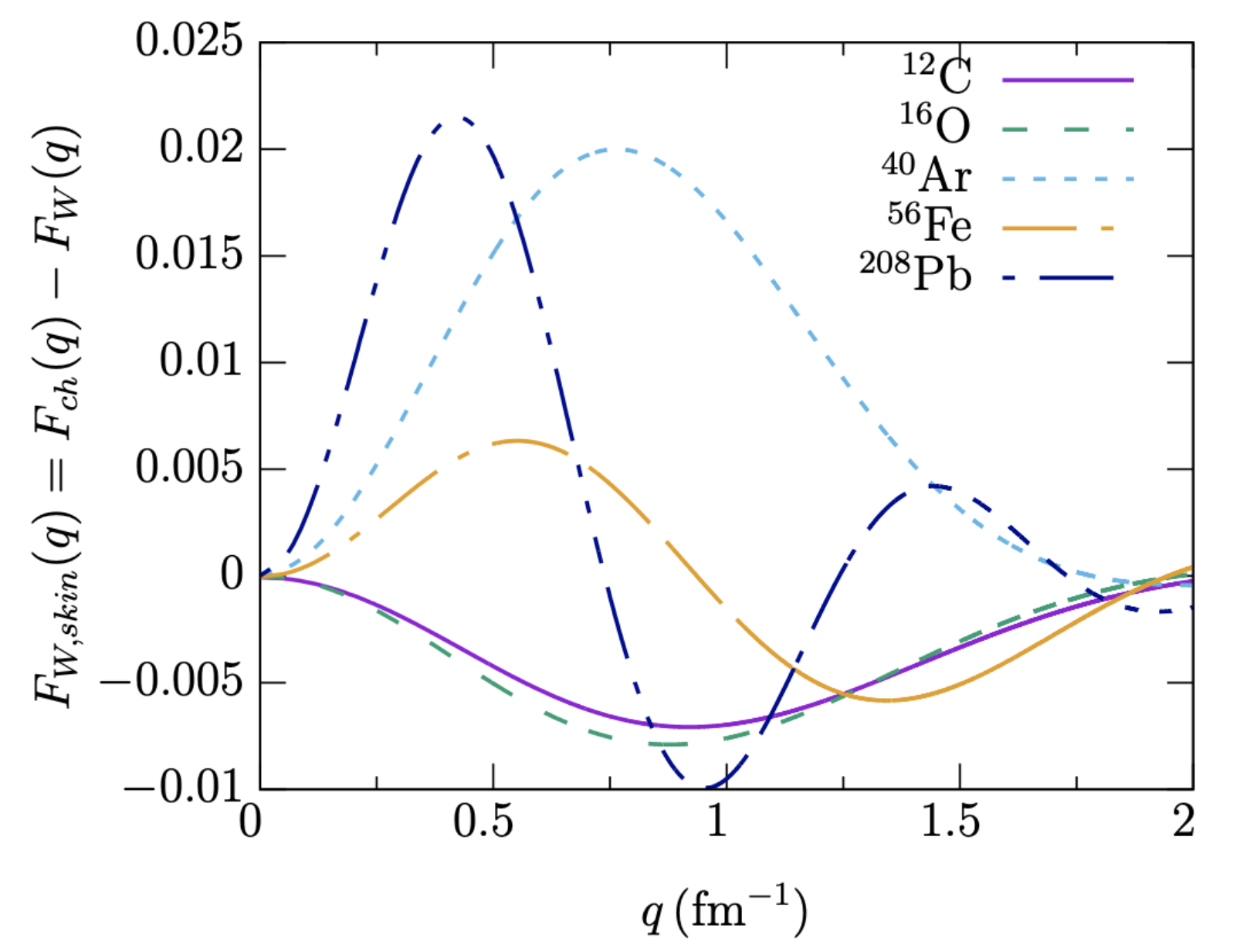}
\caption{(color online) The ``weak-skin" form factor depicts the difference between the charge and weak form factors.}
\label{Fig:FF_wskin}
\end{figure}

It is worth mentioning that the effect of long-range correlations included through the CRPA approach, vital for inelastic calculations at low energies, are found to be negligible in evaluating ground state densities of nuclei~\cite{Tohyama:2014} and are therefore not included in the elastic scattering calculations discussed in Sec.~\ref{Subsec:CEvNS_CS}.

The HF-CRPA framework offers an elegant formalism that accounts for collective excitations in the continuous spectrum as well as describes quasielastic neutrino--nucleus scattering in the low and medium energy regime. Our model has been developed over decades and has been utilized extensively to calculate various electron- and neutrino-nucleus cross sections suited for astrophysical processes as well as accelerator-based neutrino oscillation experiments~\cite{Ryckebusch:1988, Ryckebusch:1989, Jachowicz:1999, Jachowicz:2002, Jachowicz:2002_2, Jachowicz:2004, Jachowicz:2006, Pandey:2014, Pandey:2015, Pandey:2016, VanDessel:2018, VanDessel:2019, Nikolakopoulos:2019, VanDessel:2019_2, Nikolakopoulos:2020}.


\section{Results and Discussion}
\label{Sec:Results}

\begin{figure*}
\centering
\includegraphics[width=0.49\textwidth]{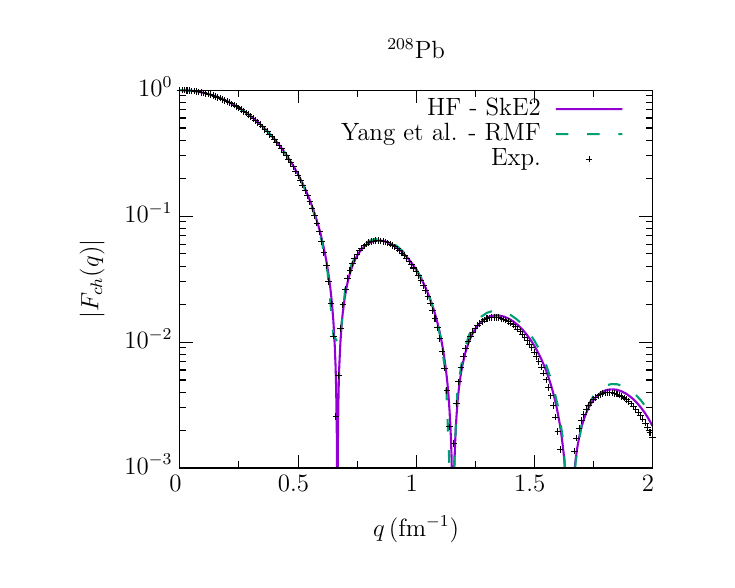}
\includegraphics[width=0.49\textwidth]{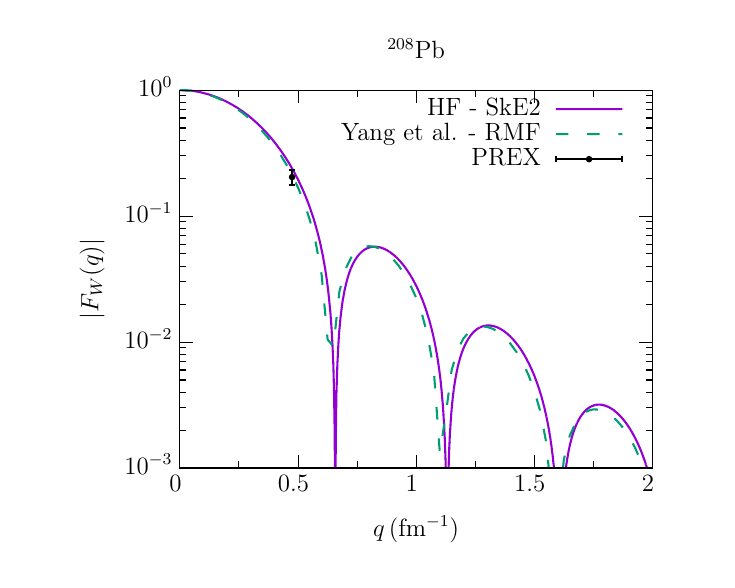}
\caption{(color online) Left: the charge form factor of $^{208}$Pb compared with elastic electron scattering data of Ref.~\cite{Vries:1987}. Right: the weak form factor of $^{208}$Pb along with the single point measured by the PREX collaboration at the momentum transfer of $q$ = 0.475~fm$^{-1}$~\cite{PREX:2012, Horowitz:2012}. Both form factors are compared with relativistic mean--field predictions of Yang {\it et al.}~\cite{Yang:2019}.}
\label{Fig:Pb_fchfw}
\end{figure*}

\par Since the weak charge of the proton is strongly suppressed by the weak mixing angle (Eq.~(\ref{Eq:cevnsformfactor2})) the nuclear weak charge is predominately carried by the neutrons. The weak form factor $F_W(Q^2)$, and hence the CE$\nu$NS cross section, are both dominated by the distribution of neutrons within the nucleus. As proton densities are well--constrained by experimental elastic electron scattering data~\cite{Angeli:2013} while little reliable neutron density data is available, phenomenological approaches approximate $\rho_{n}(r) \approx \rho_{p}(r)$ and thus assume $F_{n}(Q^2) \approx F_{p}(Q^2)$, making the nuclear form factor more of a global factor~\cite{Papoulias:2019_2}. Within the HF--SkE2 approach we treat proton and neutron densities and their corresponding form factors separately and do not have to rely on such assumptions. The densities are defined in terms of the reduced radial single particle wave functions as
\begin{equation}
    \rho_q(r) = \frac{1}{4\pi r^2} \sum_a v_{a,q}^2 (2j_a+1) |\phi_{a,q}(r)|^2,
\end{equation}
with $v_{a,q}^2$ being the occupation probability of orbital $a$ of nature $q$ (i.e. proton $p$ or neutron $n$.). In Table~\ref{table:arlevels}, we show shells and single–particle energy levels, in the case of $^{40}$Ar nucleus, as yielded by a HF calculation using SkE2 potential.

\begin{figure}
\centering
\includegraphics[width=0.8\columnwidth]{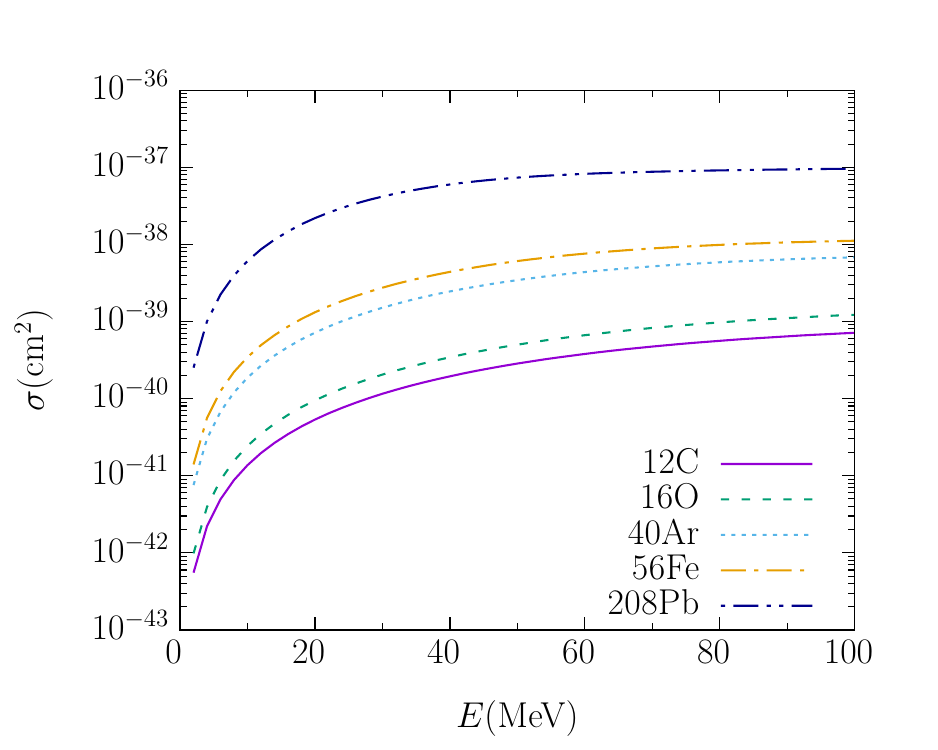}
\caption{(color online) Total CE$\nu$NS cross sections for a set of nuclear targets obtained within the HF--SkE2 approach.}
\label{Fig:coh_A}
\end{figure}

\par In Fig.~\ref{Fig:densandform}, we present proton (panel (a)) and neutron (panel (b)) density distributions of $^{12}$C, $^{16}$O, $^{40}$Ar, $^{56}$Fe and $^{208}$Pb obtained using our HF--SkE2 approach. Naturally the heavier the nucleus, the more broadly the densities are distributed. Panel (c) and (d) show the charge and weak form factors for all the nuclei. In both the charge and weak form factor cases, the heavier the nuclei the faster the form factor encounters its first minimum at rising $q$ values. Lighter nuclei have their minima spread over a larger $q$ range. $^{12}$C has its first minimum at $q \sim$ 1.8 fm$^{-1}$ while $^{208}$Pb has its first minimum around $q \sim$ 0.65 fm$^{-1}$. Although the charge and weak form factors have a similar overall structure, the minima and maxima of both occur at slightly different values of the momentum transfer, with larger differences in heavier nuclei. To further illustrate this, in Fig.~\ref{Fig:FF_wskin}, we show the ``weak--skin" form factor~\cite{Thiel:2019} for all these nuclei, defined as the difference between the charge and weak form factors:
\begin{equation}\label{Eq:fwskin}
F_{\text{W,skin}}(q) = F_{\text{ch}}(q) - F_{\text{W}}(q),
\end{equation}
which, near the origin, is proportional to the experimentally observable weak skin~\cite{Thiel:2019}. The figure illustrates that the charge and weak form factors significantly differ from each other.

\par In the left panel of Fig.~\ref{Fig:Pb_fchfw}, we show our predictions for the charge form factor of $^{208}$Pb. The predictions are compared with the experimental charge form factor obtained from a Fourier--Bessel fit to the elastic electron scattering data of Ref.~\cite{Vries:1987}. Our predictions describe the experimental data remarkably well. Our predictions almost overlap with data for $q \lesssim$ 1.8 fm$^{-1}$. We also performed a comparison with the relativistic mean--field (RMF) predictions of Yang {\it et al.}~\cite{Yang:2019}. There are no visible differences between both models up to $q \lesssim$ 1.8 fm$^{-1}$. The right panel shows our predictions for the weak form factor, again compared with the RMF predictions of~\cite{Yang:2019}. We also show the single data point measured at a momentum transfer of $q= 0.475$ fm$^{-1}$ by the PREX collaboration~\cite{PREX:2012, Horowitz:2012}. This remains the only measurement of the weak form factor obtained with an electroweak probe. The error bars on the data point are too large to discriminate between theoretical predictions. The follow--up PREX--II measurement at Jefferson lab aims to reduce the error bars by at least a factor of three. 

\begin{figure*}
\centering
\includegraphics[width=1.0\textwidth]{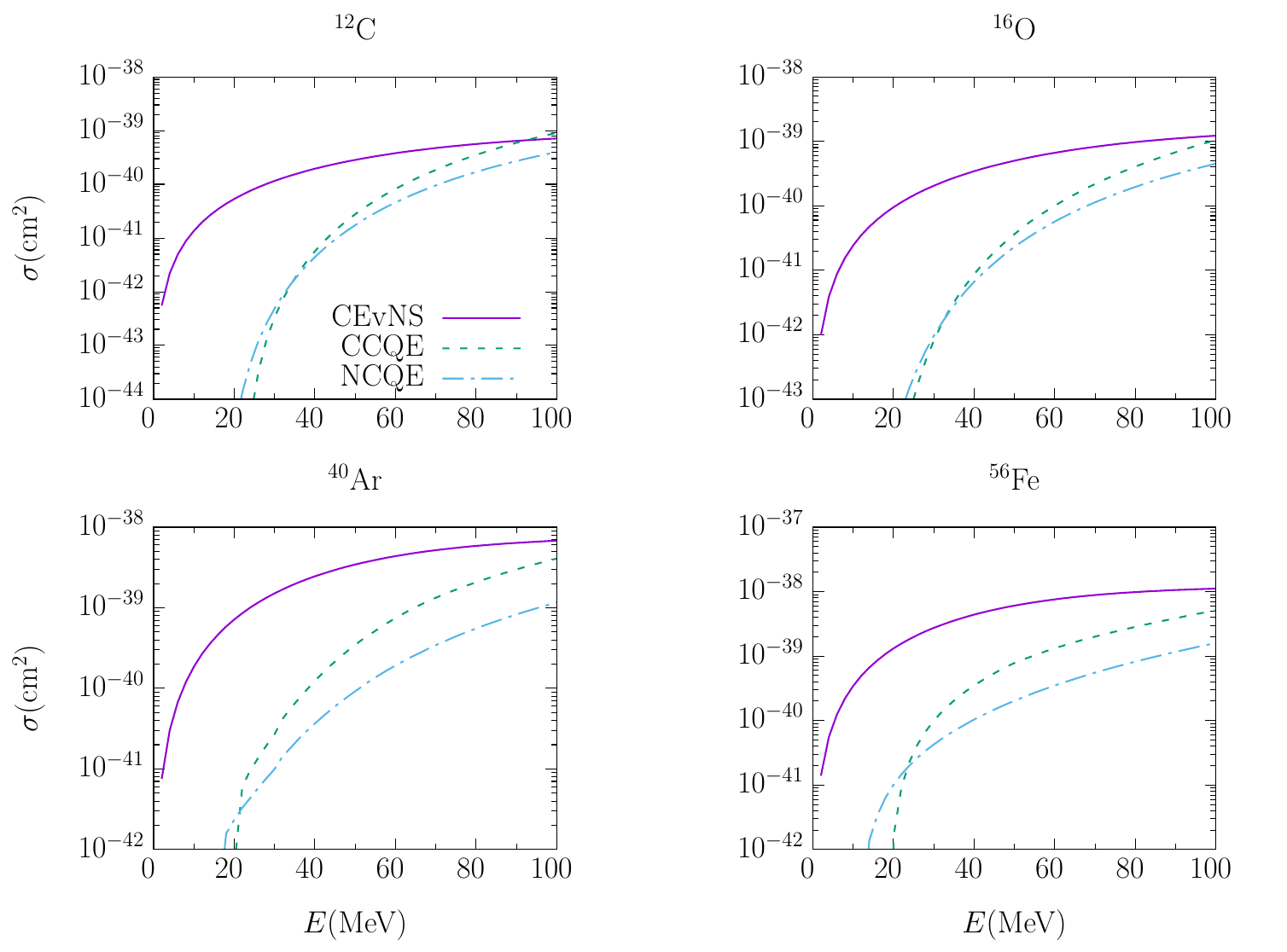}
\caption{(color online) CE$\nu$NS cross section strength compared to CCQE and NCQE scattering cross sections for several nuclei, above particle emission threshold.}
\label{Fig:cohcomp_A}
\end{figure*}

\par The total CE$\nu$NS cross section as a function of neutrino energy for $^{12}$C, $^{16}$O, $^{40}$Ar, $^{56}$Fe and $^{208}$Pb is shown in Fig.~\ref{Fig:coh_A}. All nuclei show a similar  behavior: there is a rapid rise of the cross section for incoming neutrino energies up to about $\sim$ 30 MeV, then the steep increase slows down and flattens out on the log scale thereafter. The cross section increases with the atomic number, with nearly two to three orders of magnitude difference between $^{12}$C and $^{208}$Pb, reflecting the $\approx N^2$ scaling behavior shown in Eq. \ref{eq:weakcharge}.

\par To demonstrate the dominance of the CE$\nu$NS strength over the quasi--elastic one for a neutrino energy of a few tens of MeV, in Fig.~\ref{Fig:cohcomp_A} we compare CE$\nu$NS cross sections to $\nu_e$--nucleus charged--current quasielastic (CCQE) and neutral--current quasielastic (NCQE) cross sections. 
For the energies relevant for pion decay--at--rest neutrinos, $E \lesssim 52$ MeV, the CE$\nu$NS cross section is roughly two orders of magnitude larger than inelastic cross sections.

\subsection{Constraining $^{40}$Ar} 
\label{Subsec:Argon}

\par In view of the  worldwide interest in liquid argon (LAr)--based detectors in neutrino and dark matter experiments, in this section we will focus on $^{40}$Ar. In the COHERENT collaboration's expanding series of detectors at SNS, the collaboration has recently presented new measurements from a 24 kg, single--phase, LAr CENNS--10 detector~\cite{COHERENT:2020} while a ton-scale LAr experiment is underway. A 10 ton LAr scintillation detector, Coherent CAPTAIN-Mills (CCM), was recently built at LANL to study CE$\nu$NS on $^{40}$Ar and to search for low--mass dark matter that coherently scatters off $^{40}$Ar nuclei~\cite{CCM}. Several other neutrino~\cite{SBN, DUNE} and dark matter experiments~\cite{DEAP, Darkside, Ardm, MiniClean} employ LAr detectors, making it vital to study ground state properties of the $^{40}$Ar nucleus.  

\par
\begin{figure}
\centering
\includegraphics[width=0.49\columnwidth]{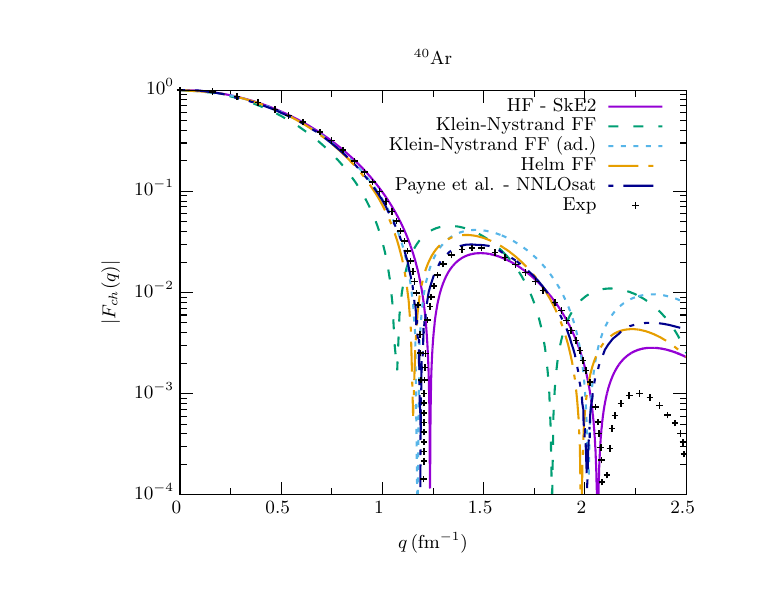}
\includegraphics[width=0.49\columnwidth]{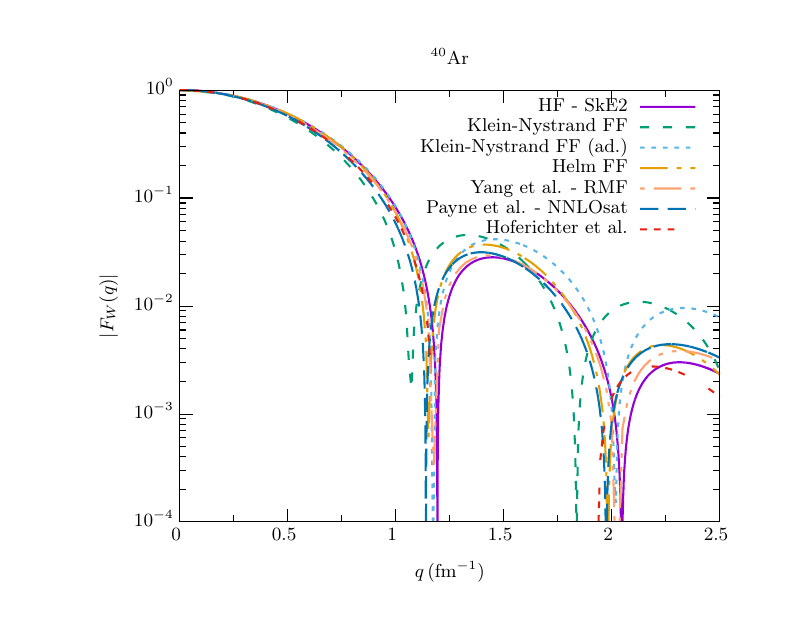}
\caption{(color online) (Left) The $^{40}$Ar charge form factor predictions compared to elastic electron scattering data taken from Ref.~\cite{Ottermann:1982}, a comparison is also performed with the coupled--cluster theory predictions of Payne {\it et al.}~\cite{Payne:2019} as well as with Klein--Nystrand~\cite{KN:1999} (standard and adapted) and Helm~\cite{Helm:1956} form factors. (Right) The $^{40}$Ar weak form factor predictions compared with calculations of Payne {\it et al.}~\cite{Payne:2019}, Yang {\it et al.}~\cite{Yang:2019}, Hoferichter {\it et al.}~\cite{Hoferichter:2020} and with the predictions of Klein--Nystrand~\cite{KN:1999} (standard and adapted) and Helm~\cite{Helm:1956} form factors.}
\label{Fig:arf}
\end{figure}

 In Fig.~\ref{Fig:arf} (left) we compare our argon charge form factor ($F_{ch}(q)$) predictions with the elastic electron scattering data of Ref.~\cite{Ottermann:1982}. Our predictions describe experimental data remarkably well for $q \lesssim 2$ fm$^{-1}$, validating our approach. We also compare with the predictions of Payne {\it et al.}~\cite{Payne:2019} where form factors are calculated within a coupled--cluster approach, using a chiral NNLO$_{\text {sat}}$ interaction. At higher $q$, $q \gtrsim 2$ fm$^{-1}$, both predictions diverge from experimental data. Note that for neutrino energies relevant for pion decay--at--rest the region above $q \gtrsim 0.5$ fm$^{-1}$ does not contribute to CE$\nu$NS cross sections. We also show a comparison with two phenomenological form factors which are widely used in the CE$\nu$NS community: the Klein--Nystrand~\cite{KN:1999} form factor that is adapted by the COHERENT collaboration and the Helm form factor~\cite{Helm:1956}. Note that we also show an adapted version of the Klein--Nystrand form factor that will be described in more detail in our discussion of form factor predictions, later in this section. 

\par After validating our approach, we make predictions for the weak form factor of $^{40}$Ar in Fig.~\ref{Fig:arf} (right). There is no data available for the weak form factor on argon yet. We compare our results with the prediction of Payne {\it et al.}~\cite{Payne:2019}, Yang {\it et al.}~\cite{Yang:2019} and Hoferichter {\it et al.}~\cite{Hoferichter:2020} as well as with the Helm form factor~\cite{Helm:1956}, the Klein--Nystrand~\cite{KN:1999} and an adapted version of the Klein--Nystrand form factor. Overall, the shape and structure of the weak form factor is similar to the charged one, but the positions of minima and maxima are somewhat different. In our HF--SKE2 approach the first minimum of $F_{\text ch}(q)$ is at $q \sim 1.23$ fm$^{-1}$ while for $F_{\text W}(q)$  it lies at $q \sim 1.19$ fm$^{-1}$, pointing to the fact that the neutron distribution extends further out compared to the proton one. To quantify differences between the charge and weak form factor, in Fig.~\ref{Fig:FF_wskin} we show the ``weak-skin" form factor of $^{40}$Ar using Eq.~(\ref{Eq:fwskin}).

\par
\begin{figure}
\centering
\includegraphics[width=1.0\columnwidth]{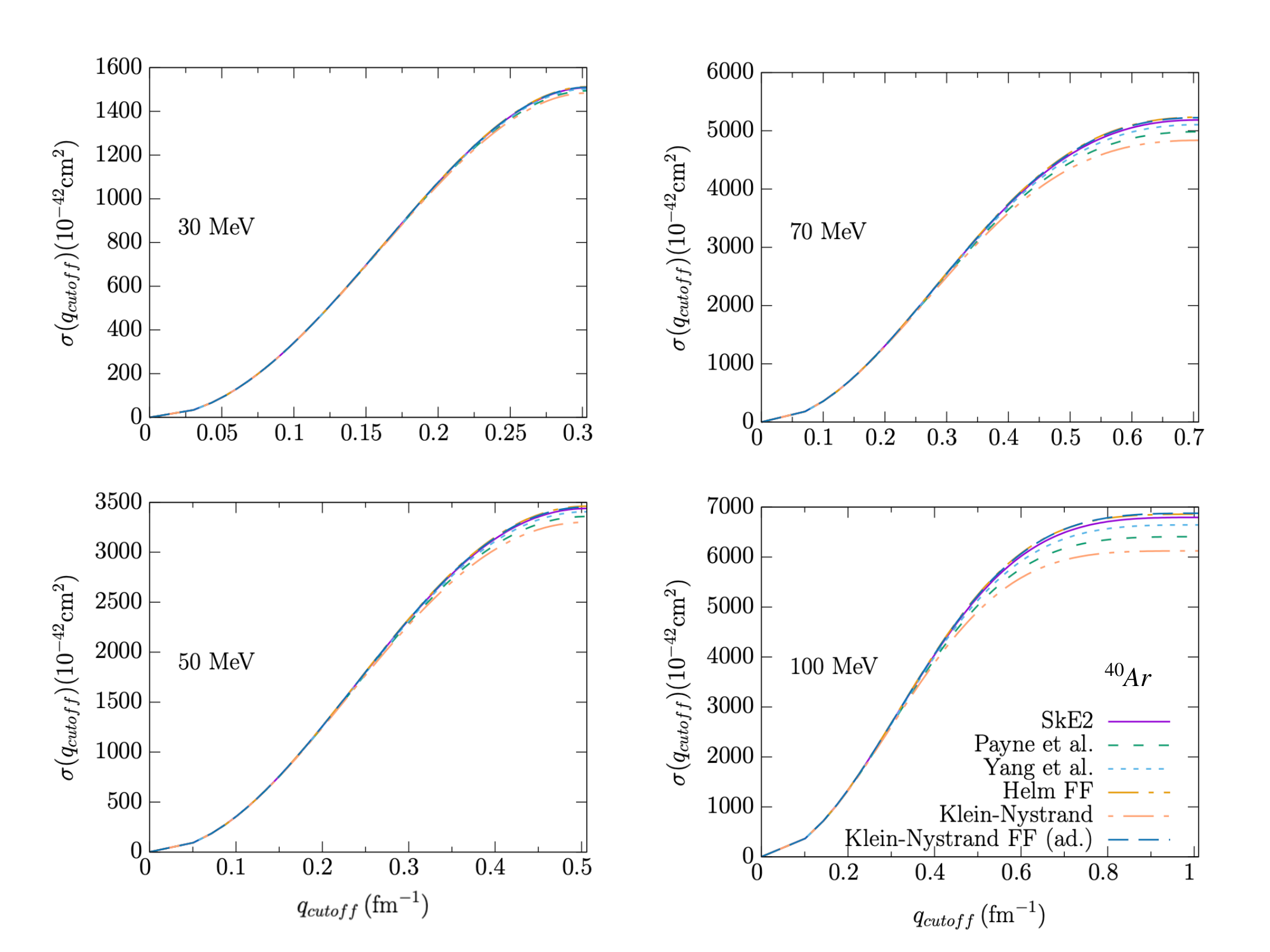}
\caption{(color online) The $^{40}$Ar cumulative cross section as a function of $q_{cutoff}$ compared with calculations done using Payne {\it et al.}~\cite{Payne:2019}, Yang {\it et al.}~\cite{Yang:2019}, as well the Klein--Nystrand~\cite{KN:1999} (standard and adapted) and Helm~\cite{Helm:1956} form factors.}
\label{Fig:arcumul}
\end{figure}

 In order to appreciate which values of momentum transfer $q$ are involved at different neutrino energies, as well as to see at which $q$ values the differences in the nuclear modeling start causing discrepancies in reaction strength predictions, we plot cumulative cross sections for $^{40}$Ar at several neutrino energies and for different models in Fig.~\ref{Fig:arcumul}. This is defined as the total cross section strength, integrated up to a cutoff value $q_{cutoff}$ in the momentum transfer:
\begin{equation}
    \sigma(q_{cutoff}) = \int_{0}^{q_{cutoff}} \frac{\mathrm{d} \sigma (q)}{\mathrm{d} q}\, \mathrm{d}q
\end{equation}
The model differences become stronger for increasingly high energies with discrepancies originating from the higher--$q$ regions of the elastic form factor. The range of cutoff values also coincides with all kinematically available momentum transfers. At 100 MeV e.g., $^{40}$Ar is only probed up to $q \approx 1 \mathrm{fm}^{-1}$.

\begin{figure*}
\centering
\includegraphics[width=1.0\textwidth]{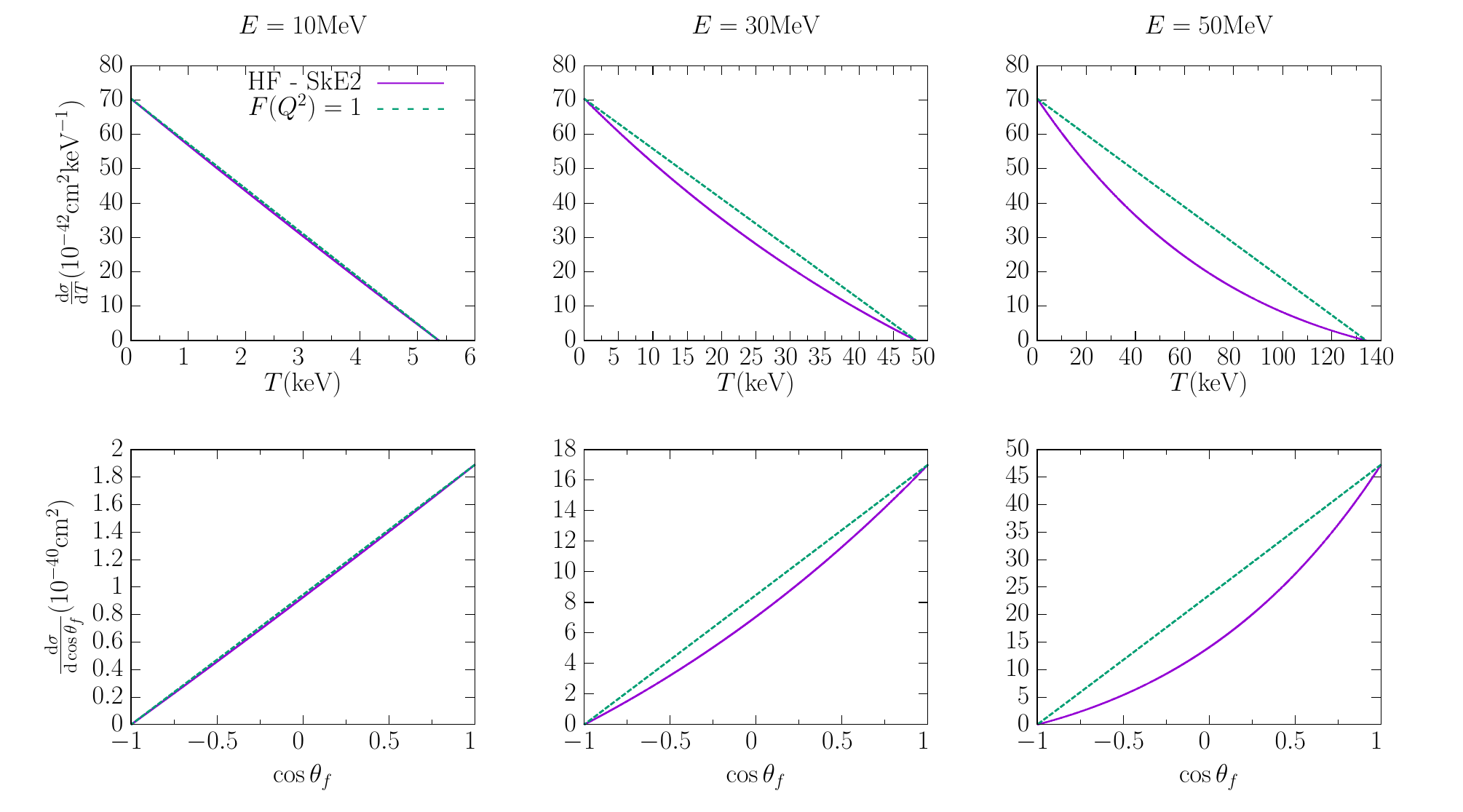}
\caption{(color online) Differential cross section on argon as a function of recoil energy and scattering angle.}
\label{Fig:ardiffcsec}
\end{figure*}
\par In Fig.~\ref{Fig:ardiffcsec}, we show differential cross sections on $^{40}$Ar as a function of recoil energy $T$, and scattering angle $\theta_f$, for different incoming neutrino energies according to Eq.~(\ref{Eq:xs_angular}) and~(\ref{Eq:xs_recoil}). For comparison, we have also plot the case with no nuclear structure effects i.e. $F(Q^2) = 1$. The effects of nuclear structure physics are more prominent as the neutrino energy increases. Most of the cross section strength lies at the lower--end of the recoil energy spectrum and for forward scattering as the cross section falls off rapidly at higher $T$ (top panels) and higher $\theta_f$ values (bottom panels). Most CE$\nu$NS detectors are sensitive only to the recoil energy deposited in the detector but, in principle, in the future more advanced detector technologies might enable measurement of both nuclear recoil and angular distribution simultaneously. Utilizing such additional information can be valuable in disentangling new physics signals in CE$\nu$NS experiments~\cite{Abdullah:2020}.

\par In Fig.~\ref{Fig:arf} (right), we come back to the differences between various predictions. Different form factor approaches are based on different representations of the nuclear densities, with no experimental data to constrain neutron distributions. Identifying the size of the differences between various theoretical predictions is crucial as experiments have to assign any deviation from expected event rates either to new physics or to unconstrained nuclear physics. We compare six predictions. These include four nuclear theory approaches: the HF--SkE2 calculation of this work, the predictions of Payne {\it et al.}~\cite{Payne:2019}, and the RMF calculations of Yang {\it et al.}~~\cite{Yang:2019} where form factors predictions are informed by  properties of finite nuclei and neutron star matter, and the predictions of Hoferichter {\it et al.}~\cite{Hoferichter:2020} where form factors are calculated using a large--scale nuclear shell model. They also contain two phenomenological approaches: the Helm~\cite{Helm:1956} and Klein--Nystrand~\cite{KN:1999} form factors where density distributions are represented by analytical expressions. 

\par In the Helm approach~\cite{Helm:1956} the density distribution is described as a convolution of a uniform nucleonic density with a given radius and a Gaussian profile characterized by the folding width $s$, accounting for the nuclear skin thickness. The resulting form factor is expressed as:
\begin{equation}
F_{\text{Helm}}(q^2) = \frac{3 j_1(qR_0)}{qR_0} e^{-q^2s^2/2},
\end{equation}
where $j_1(x) = \sin(x)/x^2 - \cos(x)/x$ is a spherical Bessel function of the first kind. $R_0$ is an effective nuclear radius given as: $R_0^2 = (1.23 A^{1/3} - 0.6)^2 + \frac{7}{3} \pi^2 r_0^2 - 5 s^2$ with $r_0$ = 0.52 fm and $s$ = 0.9 fm, fitted~\cite{Duda:2006, Lewin:1995} to muon spectroscopy and electron scattering data compiled in~\cite{Fricke:1995}. The Klein--Nystrand (KN) form factor, adapted by the COHERENT Collaboration, is obtained from the convolution of a short--range Yukawa potential with range $a_k$ = 0.7 fm over a Woods--Saxon distribution approximated as a hard sphere with radius $R_A = 1.23 A^{1/3}$ fm~\cite{KN:1999}. The resulting form factor is expressed as:
\begin{equation}
F_{\text{KN}}(q^2) = \frac{3 j_1(qR_A)}{qR_A} \left[\frac{1}{1+q^2a_k^2} \right].
\end{equation}
An adapted version of the KN form factor is often used, where $R_A$ is defined as $R_A = \sqrt{\frac{5}{3}r_0^2 - 10 a_k^2}$ utilizing measured proton rms radii $r_0$ of the nucleus~\cite{Aristizabal:2019, Papoulias:2019_2}. We show both the standard and the adapted (ad.) KN form factor. For the adapted one we use $r_0 = 3.427$ fm, the measured proton rms radii of $^{40}$Ar~\cite{Angeli:2013}.

\par We attempt to quantify differences between different form factors and the CE$\nu$NS cross section due to different underlying nuclear structure details. We consider quantities that emphasize the relative differences between the results of different calculations, arbitrarily using HF--SkE2 as a reference calculation, as follows:
\begin{equation}
|\Delta F_{\text W}^{i}(q)|~=~ \frac{|F_{\text W}^{i}(q) - F_{\text W}^{\text {HF}}(q)|}{|F_{\text W}^{\text {HF}}(q)|},
\end{equation}
and
\begin{equation}
\Delta \sigma_{\text W}^i(E)~=~ \frac{|\sigma_{\text W}^{i}(E) - \sigma_{\text W}^{\text {HF}}(E)|}{\sigma_{\text W}^{\text {HF}}(E)},
\end{equation}
where $i$ refers to calculations from different approaches as discussed above.

\begin{figure*}
\centering
\includegraphics[width=1.0\textwidth]{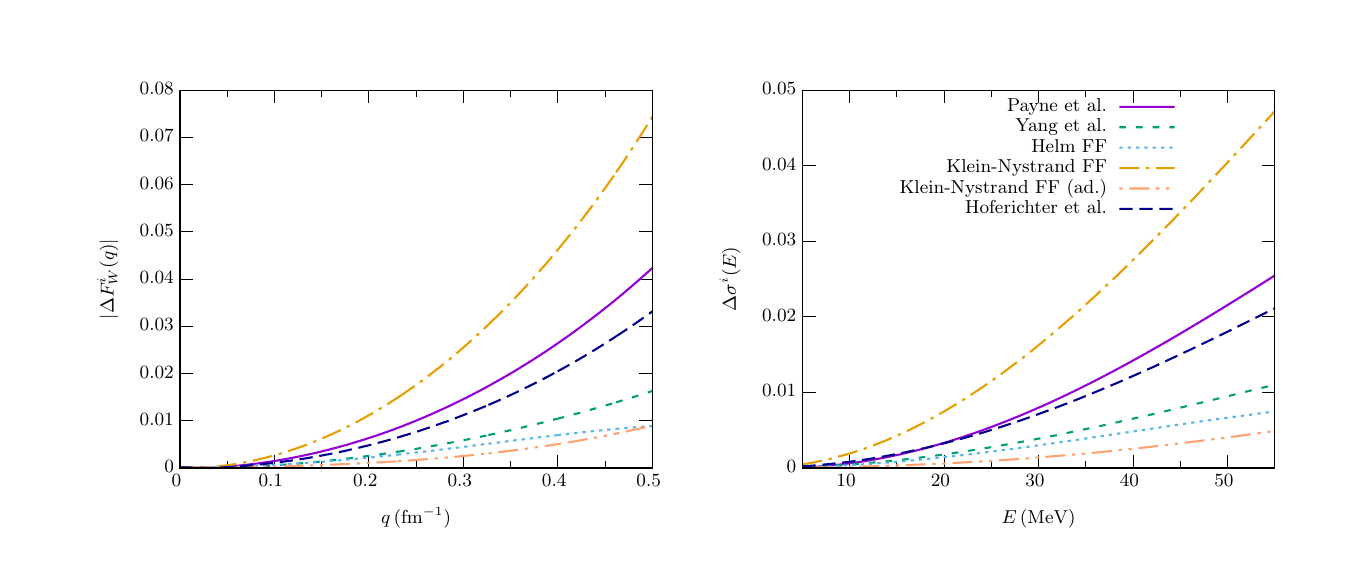}
\caption{(color online) Relative differences in the weak form factor and CE$\nu$NS cross section predictions of Payne {\it et al.}~\cite{Payne:2019}, Yang {\it et al.}~\cite{Yang:2019}, Hoferichter {\it et al.}~\cite{Hoferichter:2020}, Helm~\cite{Helm:1956}, Klein--Nystrand~\cite{KN:1999} and the adapted Klein--Nystrand~\cite{Aristizabal:2019, Papoulias:2019_2}, all with respect to HF-SkE2.}
\label{Fig:arcomps}
\end{figure*}
\begin{figure}
\centering
\includegraphics[width=0.9\columnwidth]{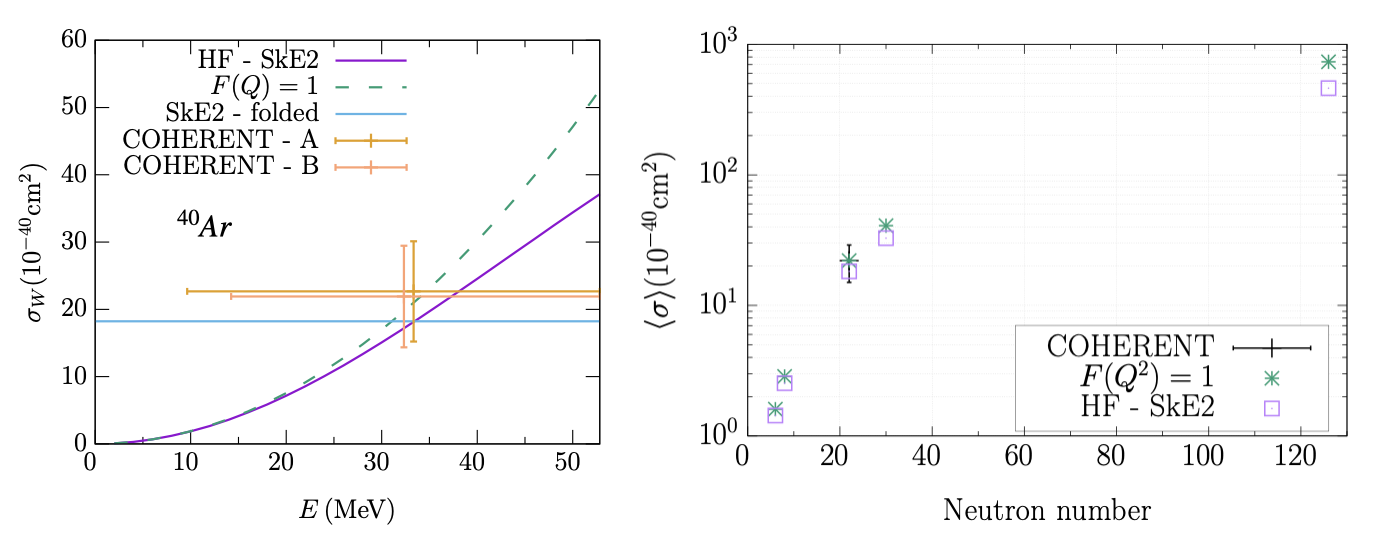}
\caption{(color online) (Left) The CE$\nu$NS cross section on $^{40}$Ar as a function of neutrino energy, recent flux--folded measurement by the COHERENT collaboration~\cite{COHERENT:2020} is shown along with the flux-folded HF--SkE2 prediction. (Right) Flux--averaged CE$\nu$NS cross sections as a function of neutron number for the $^{12}$C, $^{16}$O, $^{40}$Ar, $^{56}$Fe and $^{208}$Pb nuclei. We also show $^{40}$Ar data measured by COHERENT~\cite{COHERENT:2020}.}
\label{Fig:arcseccoherentcomp}
\end{figure}
\begin{figure}
\centering
\includegraphics[width=0.49\columnwidth]{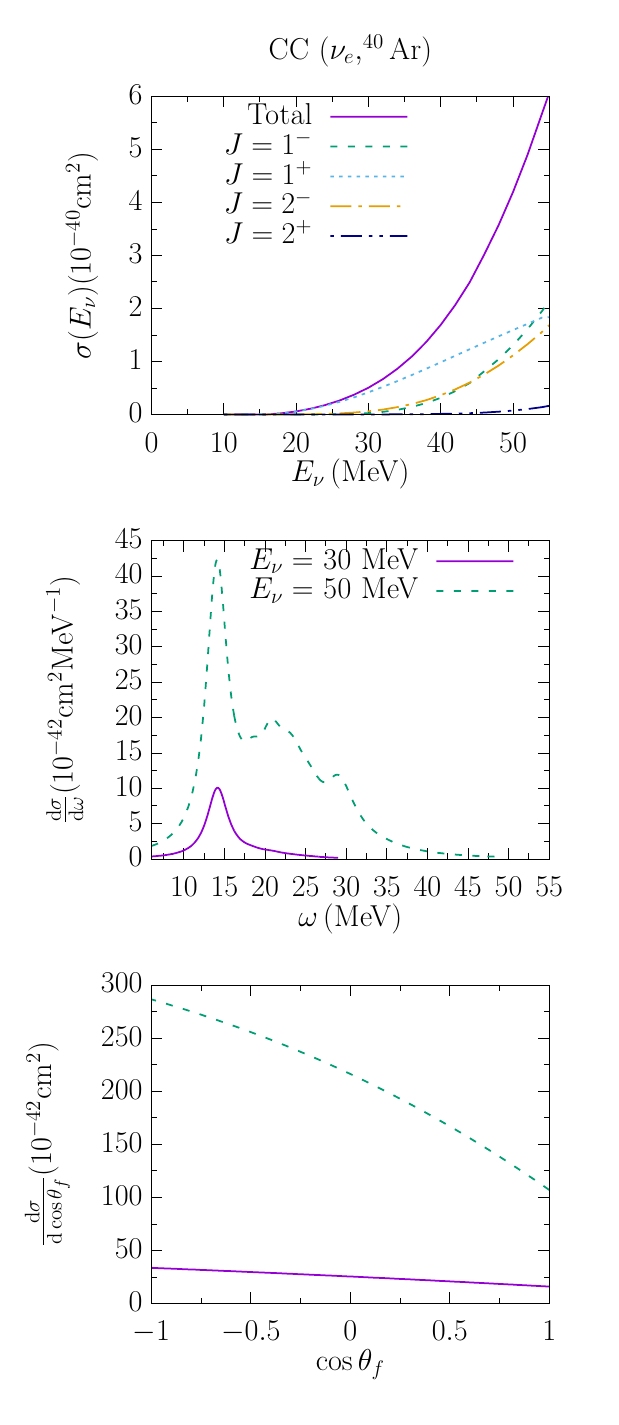}
\includegraphics[width=0.49\columnwidth]{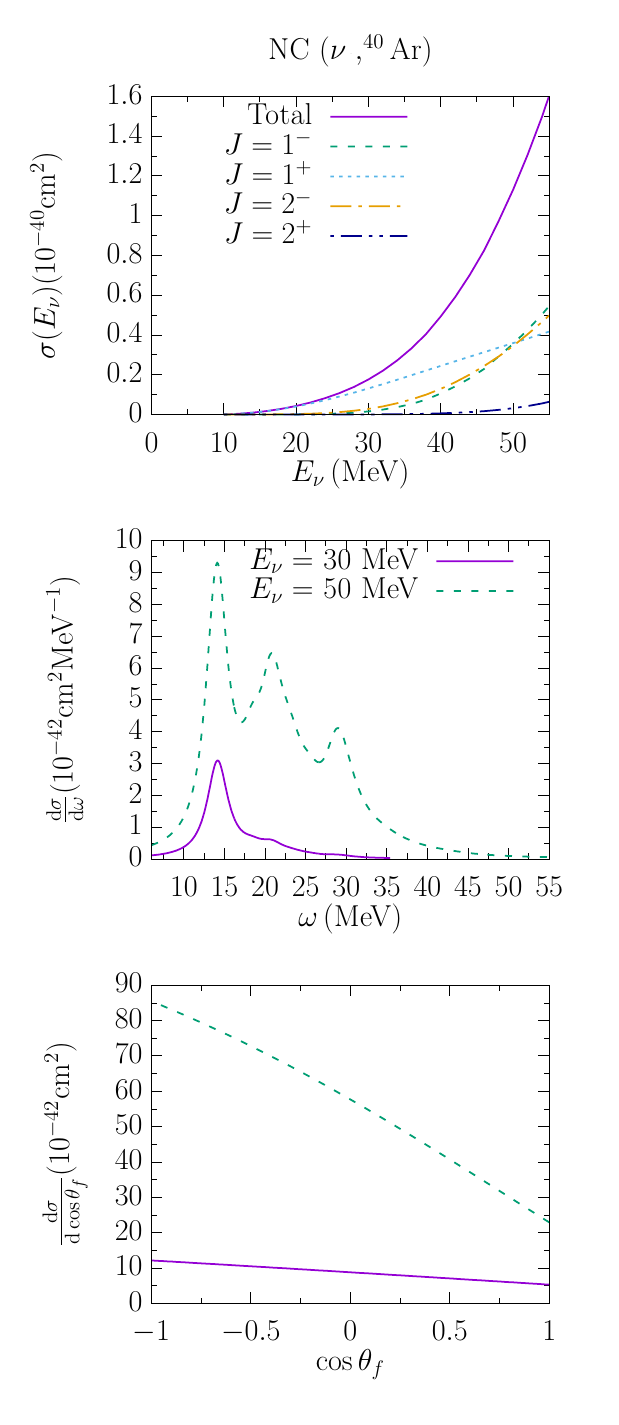}
\caption{(color online) Charged-current (left) and neutral-current (right) inelastic cross section: total as a function of neutrino energy shown along with contributions from different multipoles (top panel), differential as a function of excitation energy (middle panel) and as a function of lepton scattering angle (bottom panel) for fixed neutrino energies, $E_\nu =$ 30 and 50 MeV.}
\label{Fig:CC_NC_Ar}
\end{figure}

\par The relative differences are shown in Fig.~\ref{Fig:arcomps}. We show only the low--momentum part of the weak form factor to a maximum value of $q$ = 0.5 fm$^{-1}$ ($\sim$ 100 MeV) that corresponds to a maximum incoming neutrino energy of E $\sim$ 50 MeV, as shown in Fig.~\ref{Fig:arcumul}. The relative differences are shown on a linear scale. At smaller energies the momentum transfer is low and hence the differences between form factors are also small. For higher energies the available momentum transfer increases and therefore the differences between the form factors become more prevalent. The differences in model predictions amount to $< 7.5\%$ over the entire momentum transfer range. The differences rise rapidly at the higher end of the $q$ range. This translates into relative differences in CE$\nu$NS cross sections, $\Delta \sigma(E)$, of $< 5\%$ over the whole energy range, where $E \lesssim 55$ MeV, relevant for neutrinos from pion decay-at-rest. Note that most of the strength in the cross section lies at the lower $T$ end (and therefore at the lower $q$ end), as we have seen in Fig.~\ref{Fig:ardiffcsec}.

\par The CE$\nu$NS cross section on $^{40}$Ar as a function of the neutrino energy is shown in Fig.~\ref{Fig:arcseccoherentcomp} (left). We also show recent flux--averaged measurements performed by the COHERENT collaboration~\cite{COHERENT:2020}. Measurements from two analyses are included, with the horizontal bars indicating the minimum value set by the nuclear recoil threshold energy for each analysis. The flux--averaged measured cross section is 2.2 $\pm$ 0.7 $\times$ 10$^{-39}$cm$^2$ (average of both analyses), while the HF-SkE2 predicted flux--averaged cross section is 1.82 $\times$ 10$^{-39}$cm$^2$. The total experimental error is dominated by statistics, amounting to $\sim$ 30$\%$. Future measurements by ton--scale LAr detector at SNS and 10--ton LAr detector CCM at LANL will be able to provide more precise measurements of the CEvNS cross section on $^{40}$Ar. In Fig.~\ref{Fig:arcseccoherentcomp} (right), we also show flux--folded cross sections as a function of neutron number for all five nuclei -- $^{12}$C, $^{16}$O, $^{40}$Ar, $^{56}$Fe and $^{208}$Pb -- considered in this paper. As expected, the deviation of $F(Q^2) = 1$ from the full HF-SkE2 calculation becomes more prominent as the number of neutrons, and hence the influence of nuclear structure effects, increases. Also included is the $^{40}$Ar data measured by COHERENT~\cite{COHERENT:2020}.

\par CE$\nu$NS liquid argon detectors at stopped--pion sources are well suited to measure inelastic cross sections as well. Inelastic cross section measurements on $^{40}$Ar will provide powerful constraints on supernova detection capabilities of future kiloton liquid argon experiments such as DUNE~\cite{DUNE}. In view of this, in Fig.~\ref{Fig:CC_NC_Ar} we present CC (left) inelastic $(\nu_e, ^{40}$Ar$)$ and NC (right) inelastic $(\nu, ^{40}$Ar$)$ cross sections for energies relevant to pion decay--at--rest neutrinos. These cross sections are calculated by incorporating the CRPA approach on top of the initial HF--SkE2  nuclear picture. 

\par The top panels in Fig.~\ref{Fig:CC_NC_Ar} show total cross section as a function of incoming neutrino energy along with separate contributions coming from the dominating individual multipoles. In both CC and NC case, most strength arises from $1^-$, $1^+$ and $2^-$ multipoles. The $0^+$ and $0^-$ transitions contribute only minimally to the total reaction strength  for excitations into the continuum and are not shown here. Still, it is clear that a considerable part of the strength stems from forbidden transitions.  The middle panels show the differential cross sections as a function of excitation energy $\omega$ for two incoming neutrino energies $E_\nu$ = 30 MeV and 50 MeV. As the energy increases, more resonance peaks show up as an increasing number of excitations becomes accessible. Differential cross sections are folded with a Lorentzian of width 3 MeV in order to account for the finite width of the resonances~\cite{Pandey:2015}. The bottom panels show the differential cross sections as a function of the direction of the outgoing lepton scattering angle $\cos\theta_f$ for two incoming neutrino energies $E_\nu$ = 30 MeV and 50 MeV. The differential cross sections in scattering angles favor backward scattering.

\section{Conclusions}
\label{Sec:Conclusions}

\par The experimental observation of coherent elastic neutrino--nucleus scattering processes by the COHERENT collaboration has inspired physicists across many fields. The power of CE$\nu$NS as a probe  of BSM physics and its potential for determining neutron density distributions is becoming more and more apparent. The main uncertainty in the evaluation of the CE$\nu$NS cross sections is driven by the weak form factor that encodes the entire nuclear structure contribution to the CE$\nu$NS cross section. 

\par We presented microscopic nuclear physics calculations of charge and weak nuclear form factors and the CE$\nu$NS cross section on $^{12}$C, $^{16}$O, $^{40}$Ar, $^{56}$Fe and $^{208}$Pb nuclei. We obtain neutron (proton) densities and weak (charge) form factors by solving the Hartree--Fock equations with a Skyrme (SkE2) nuclear potential. Our predictions for $^{208}$Pb and $^{40}$Ar charge form factors describe elastic electron scattering data remarkably well. 

\par After validating $^{40}$Ar charge form factor calculations, we make predictions for the $^{40}$Ar weak form factor. Thereby, we calculate differential cross section as a function of recoil energy and neutrino scattering angle. We attempt to gauge the level of theoretical uncertainty pertaining to the description of $^{40}$Ar form factor and CE$\nu$NS cross section by comparing relative differences between recent nuclear theory and widely--used phenomenological form factor predictions. We compare our $^{40}$Ar prediction with recent measurements of the COHERENT collaboration. Future precise measurements of CE$\nu$NS with ton and multi--ton detectors will aid in constraining neutron densities and weak nuclear form factor that will in turn improve prospects of extracting new physics through CE$\nu$NS.

Furthermore, we calculate inelastic charged--current and neutral--current cross section on $^{40}$Ar within the same formalism, 
and comparing the strength of coherent and inelastic processes.  We present total and differential cross sections as a function of excitation energy and lepton scattering angle for neutrino energy relevant for pion decay--at--rest neutrinos. CE$\nu$NS experiments at stopped--pion sources are well-suited to measure these inelastic cross sections and can provide powerful constraints on supernova detection capabilities of future kiloton liquid argon experiments.


\acknowledgments 
We thank S.~Bacca, J.~Yang and M.~Hoferichter for providing their calculations for comparison. We thank Richard Van de Water for fruitful discussions. NVD and NJ are supported by the Research Foundation Flanders (FWO–Flanders). VP and HR acknowledge the support from US DOE under grant DE-SC0009824. This manuscript has been authored by Fermi Research Alliance, LLC under Contract No. DE-AC02-07CH11359 with the U.S. Department of Energy, Office of Science, Office of High Energy Physics.


\end{document}